\documentclass[a4paper,twocolumn,11pt,accepted=2020-05-13]{quantumarticle}
\pdfoutput=1
\usepackage[utf8]{inputenc}
\usepackage[english]{babel}
\usepackage[T1]{fontenc}
\usepackage{amsmath}
\usepackage{hyperref}

\usepackage{tikz}
\usepackage{lipsum}

\usepackage{color}
\usepackage[]{graphicx}
\usepackage{amsfonts}
\usepackage[numbers,sort&compress]{natbib}

\def\ket#1{| #1 \rangle}
\def\bra#1{\langle #1 |}
\def\bracket#1#2{\langle #1 | #2 \rangle}
\def\kb#1#2{| #1 \rangle\!\langle #2 |}

\def\cE{\mathcal{E}}

\def\cM{\mathcal{M}}
\def\cN{\mathcal{N}}
\def\cO{\mathcal{O}}

\def\cW{\mathcal{W}}
\def\cX{\mathcal{X}}

\def\eq#1{Eq.~\eqref{eq:#1}}
\def\fig#1{Fig.~\ref{fig:#1}}
\def\eq#1{Eq.~\eqref{eq:#1}}
\def\Sec#1{Sec.~\ref{sec:#1}}

\begin{document}

\title{Efficient Quantum Walk Circuits for Metropolis-Hastings Algorithm}

\author{Jessica Lemieux}
\affiliation{D\'epartement de Physique \& Institut Quantique, Universit\'e de Sherbrooke, Qu\'ebec, Canada}
%\orcid{0000-0002-2445-2701}
\author{Bettina Heim}
\affiliation{Quantum Architecture and Computation Group, Microsoft Research, Redmond, WA 98052, USA}
\author{David Poulin}
%\email{David.Poulin@USherbrooke.ca}
\affiliation{D\'epartement de Physique \& Institut Quantique, Universit\'e de Sherbrooke, Qu\'ebec, Canada}
\affiliation{Canadian Institute for Advanced Research, Toronto, Ontario, Canada M5G 1Z8}
\author{Krysta Svore}
\affiliation{Quantum Architecture and Computation Group, Microsoft Research, Redmond, WA 98052, USA}
\author{Matthias Troyer}
\affiliation{Quantum Architecture and Computation Group, Microsoft Research, Redmond, WA 98052, USA}
%\thanks{You can use the \texttt{\textbackslash{}email}, \texttt{\textbackslash{}homepage}, and \texttt{\textbackslash{}thanks} commands to add additional information for the preceding \texttt{\textbackslash{}author}. If applicable, this can also be used to indicate that a work has previously been published in conference proceedings.}

\maketitle

\begin{abstract}
  We present a detailed circuit implementation of Szegedy's quantization of the Metropolis-Hastings walk. This quantum walk is usually defined with respect to an oracle. We find that a direct implementation of this oracle requires costly arithmetic operations. We thus reformulate the quantum walk, circumventing its implementation altogether by closely following the classical Metropolis-Hastings walk. We also present heuristic quantum algorithms that use the quantum walk in the context of discrete optimization problems and numerically study their performances. Our numerical results indicate polynomial quantum speedups  in heuristic settings. 
\end{abstract}

Markov chain Monte Carlo (MCMC) methods are a cornerstone of modern computation, with applications ranging from computational science to machine learning. The key idea is to sample a distribution $\pi_x$ by constructing a random walk $\cW$ which reaches this distribution at equilibrium $\cW\pi =\pi$. One important characteristic of a Markov chain is its mixing time, the time it requires to reach equilibrium. This mixing time is governed by the inverse spectral gap of $\cW$, where the spectral gap $\Delta$ is defined as the difference between its two largest eigenvalues. The runtime of a MCMC algorithm is thus determined by the product of the mixing time and the time required to implement a single step of the walk.

Szegedy \cite{S04a} presented a general method to quantize reversible walks, resulting in a unitary transformation $U_\cW$. The eigenvalues $e^{i\theta_j}$ of a unitary matrix all lie on the unit complex circle, and we choose $0=\theta_0\leq \theta_1\leq\theta_2\leq\ldots$ The steady state $\ket \pi $ of the quantum walk is essentially a coherent version $\ket \pi  = \sum_x \sqrt{\pi_x} \ket x$ of the classical equilibrium distribution $\pi$. The main feature of the quantum walk is that its spectral gap $\delta := \theta_1\geq \sqrt{\Delta}$ is quadratically larger than its classical counterpart. Combined with the quantum adiabatic algorithm \cite{FGGS00a,AT03b,BKS10a1}, this yields a quantum algorithm to reach the steady state that scales quadratically faster with $\Delta$ than the classical MCMC algorithm \cite{SBBK08a}. 

While at first glance this is an important advantage with far-reaching applications, additional considerations must be taken into account to determine if quantum walks offer a significant speedup for any specific application. One of the reasons is that it coud take significantly longer to implement a single step $U_\cW$ of the quantum walk than to implement a step $\cW$ of the classical walk. Thus, quantum walks are more likely to offer advantages in situations with extremely long equilibration times. Moreover, we must address the fact that classical walks are often used heuristically out of equilibrium. When training a neural network for instance, where a MCMC method called stochastic gradient descent is used to minimize a cost function, it is in practice often not necessary to reach the true minimum, and thus the MCMC runs in time less than its mixing time. Similarly,  simulated annealing is typically used heuristically with cooling schedules far faster than prescribed by provable bounds -- and combined with repeated restarts. Such heuristic applications further motivate the constructions of efficient implementations of $U_\cW$, and the development of heuristic methods for quantum computers.

This article addresses these two points. First, we present a detailed realization and cost analysis of the quantum walk operator for the special case of a Metropolis-Hastings walk \cite{MRR+53a,H70a}. This is a widespread reversible walk, whose implementation only requires knowledge of the relative populations $\pi_x/\pi_y$ of the equilibrium distribution. While Szegedy's formulation of the quantum walk builds on a classical walk oracle, our implementation circumvents its direct implementation, which would require costly arithmetic operations. Instead, we directly construct a related but different quantum unitary walk operator with an effort to minimize circuit depth. Second, we suggest heuristic uses of this oracle inspired by the adiabatic algorithm, and study their performances numerically.

\section{Preliminaries}

\subsection{Quantum Walk}
\label{sec:walk}

We define a classical walk on a $d$-dimensional state space $\cX=\{x\}$ by a $d\times d$ transition matrix $\cW$ where the transition probability $x\rightarrow y$ is given by matrix element $\cW_{yx}$. Thus, the walk maps the distribution $p$ to the distribution $p' = \cW p$, where $p'_y = \sum_x \cW_{yx}p_x$. An aperiodic walk is {\em irreducible} if  every state in $\cX$ is accessible from every other state in $\cX$, which implies the existence of a unique equilibrium distribution $\pi = \cW\pi$.  Finally, a walk is {\em reversible} if it obeys the detailed balance condition 
\begin{equation}
\cW_{yx}\pi_x = \cW_{xy}\pi_y.
\label{eq:db}
\end{equation}
We now explain how to quantize a reversible classical walk $\cW$.

Szegedy's quantum walk \cite{S04a} is  formulated in an oracle setting. For a classical walk $\cW$, it assumes a unitary transformation $W$ acting on a Hilbert space $\mathbb{C}^d \otimes \mathbb{C}^d$ with the following action
\begin{equation}
W\ket x\otimes \ket 0 = \ket{w_x}\otimes \ket x =: \ket{\phi_x},
\label{eq:W}
\end{equation}
where $\ket{w_x} := \sum_y \sqrt{\cW_{yx}}\ket y$. Define $\Pi_0$ as the projector onto the subspace $\cE_0$ spanned by states $\{\ket x\otimes \ket 0\}_{x=1}^d$. Combining $W$ to the reflection $R = 2\Pi_0 - I$ and the swap operator $\Lambda$, we can construct the quantum walk defined by 
\begin{align}
U_\cW 
&:=  RW^\dagger \Lambda W \label{eq:UW}\\ 
&=  (2\Pi_0-1) W^\dagger \Lambda W  \label{eq:WLW}.
\end{align}
Szegedy's walk is defined as $\Lambda W(RW^\dagger \Lambda W) RW^\dagger$, so it is essentially the square of the operator $U_\cW$ we have defined, but this will have no consequence on what follows aside from a minor simplification.

To analyze the quantum walk $U_\cW$, let us define the state $\ket{\psi_x} := \Lambda\ket{\phi_x} = \ket x\otimes\ket{w_x}$ and consider the operator
\begin{align}
X &:= \Pi_0 W^\dagger \Lambda W \Pi_0 \\
&= \sum_{xy}\bracket{\phi_y}{\psi_x}  \kb yx \otimes \kb 00  \\
&= \sum_{xy} \sqrt{\cW_{xy} \cW_{yx}} \kb yx \otimes \kb 00.
\end{align}
At this point, in order to use detailed balance condition of \eq{db}, we need to assume that the walk is reversible to obtain
\begin{align}
X 
&= \sum_{xy} \sqrt{\frac{\pi_x}{\pi_y} }\cW_{yx} \kb yx \otimes \kb 00,
\label{eq:X}
\end{align}
or, if we restrict the operator $X$ to its support $\cE_0$, we get in matrix notation $X = {\rm diag}(\pi^{-\frac 12}) \cW \ {\rm diag}(\pi^{\frac 12})$. The matrices $X$ and $\cW$ are thus similar so they have the same eigenvalues. Define its eigenvectors 
\begin{align}
X\ket{\tilde \gamma_k} = \lambda_k\ket{\tilde \gamma_k},
\end{align}
where $\lambda_k$ are the eigenvalues of $\cW$. Because the operator $X$  is obtained by projecting the operator $W^\dagger \Lambda W$ onto the subspace $\cE_0$, its eigenvectors with non-zero eigenvalues in the full Hilbert space must have the form $\ket{\gamma_k} = \ket{\tilde\gamma_k}\otimes \ket 0$. 

If we consider the action of $W^\dagger \Lambda W$ without those projections, we get
\begin{equation}
W^\dagger \Lambda W \ket{\gamma_k} = \lambda_k\ket{\gamma_k} - \beta_k \ket{\gamma_k^\perp}
\label{eq:WW}
\end{equation} 
where $\ket{\gamma_k^\perp}$ is orthogonal to the subspace $\cE_0$, so in particular it is orthogonal to all the vectors $\ket{\gamma_{k'}}$. Finally, because $W^\dagger \Lambda W$ is a unitary, we also obtain that the  $\ket{\gamma_k^\perp}$ are orthogonal to each other and that $\beta_k = \sqrt{1-|\lambda_k|^2}$. This implies that the vectors $\{ \ket{\gamma_k}, \ket{\gamma_k^\perp}\}$ are all mutually orthogonal and that $W^\dagger\Lambda W$ is block diagonal in that basis. 

Given the above observations, it is straightforward to verify that 
\begin{align}
U_\cW \ket{\gamma_k} &= \lambda_k\ket{\gamma_k} + \sqrt{1-|\lambda_k|^2}\ket{\gamma_k^\perp} \\
U_\cW \ket{\gamma_k^\perp} &= \sqrt{1-|\lambda_k|^2}\ket{\gamma_k}  -\lambda_k\ket{\gamma_k^\perp},
\end{align}
so the eigenvalues of $U_k$ on the subspace spanned by $\{ \ket{\gamma_k}, \ket{\gamma_k^\perp}\}$ are $e^{\pm i\theta_k}$ where $\cos\theta_k = \lambda_k$ with corresponding eigenvectors $\ket{\gamma_k^\pm} =  \frac 1{\sqrt 2}(\ket{\gamma_k}\pm i \ket{\gamma_k^\perp})$.

\subsection{Adiabatic state preparation}
\label{sec:zeno}

We can use quantum phase estimation \cite{Kit95a} to measure the eigenvalues of $U_\cW$. In particular, we want this measurement to be sufficiently accurate to resolve the eigenvalue $\theta = 0$, or equivalently  $\lambda_k = 1$, from the rest of the spectrum. Assuming that the initial state is supported on the subspace $\cE_0$, the spectral gap of $U_\cW$ is $\delta =\theta_1 = \arccos(\lambda_1) = \arccos(1-\Delta) \sim \sqrt\Delta$, so we only need about $1/\sqrt{\Delta}$ applications of $U_\cW$ to realize that measurement. This is quadratically faster than the classical mixing time $1/\Delta$, which is the origin of the quadratic quantum speed-up.

A measurement outcome corresponding to $\theta = 0$ would produce the coherent stationary distribution $\ket\pi\otimes \ket 0  := \sum_x \sqrt{\pi_x} \ket x \otimes \ket 0$. Indeed, first note that for any $\ket\psi$ such that $X(\ket\psi\otimes\ket 0) = \ket\psi\otimes\ket 0$, \eq{WW} implies that $U_\cW(\ket\psi\otimes\ket 0) = \ket\psi\otimes\ket 0$. We can verify that this condition holds for $\ket\psi = \ket\pi$:
\begin{align}
 X \sum_x \sqrt{\pi_x}\ket x\otimes \ket 0 
&= \sum_{xy} \sqrt{\frac{\pi_x}{\pi_y} }\cW_{yx} \sqrt{\pi_x} \ket y \otimes \ket 0 \\
&= \sum_{xy} \cW_{xy} \sqrt{\pi_y} \ket y \otimes \ket 0 \\
&= \sum_{y} \sqrt{\pi_y} \ket y \otimes \ket 0
\end{align}
where we have used detailed balance \eq{db} in the second step and $\sum_x \cW_{xy} = 1$ in the last step. 

From an initial state $\ket\psi\otimes \ket 0 = \sum_k \alpha_k \ket{\gamma_k}$, the probability of that measurement outcome is $|\bracket\psi\pi|^2 = |\alpha_0|^2$. Therefore, the initial state $\ket\psi$ must be chosen with a large overlap with the fixed point to ensure that this measurement outcome has a non-negligible chance of success. If no such state can be efficiently prepared, one can use adiabatic state preparation \cite{FGGS00a,AT03b} to increase the success probability. In its discrete formulation \cite{SBBK08a} inspired by the quantum Zeno effect, we can choose a sequence of random walks $\cW^0,\cW^1,\ldots \cW^L = \cW$ with coherent stationary distributions $\ket{\pi^j}$. The walks are chosen such that $\ket{\pi^0}$ is easy to prepare and consecutive walks are  nearly identical, so that $|\bracket{\pi^j}{\pi^{j+1}}|^2\geq 1-\frac 1L$~\cite{SBBK08a}. Thus, the sequence of $L$ measurements of the eigenstate of the corresponding quantum walk operators $U_{\cW^j}$ all yield the outcomes $\theta=0$ with probability $(1-\frac 1L)^L \sim \frac 1e$, which results in the desired state. The overall complexity of this algorithm is 
\begin{equation}
C \sum_{j=1}^L \frac 1{\delta_j}
\end{equation}
where $\delta_j$ is the spectral gap of the $j$-th quantized walk $\cW^j$ and $C$ is the time required to implement a single quantum walk operator.

\subsection{Metropolis-Hastings Algorithm}

The Metropolis-Hastings algorithm \cite{MRR+53a,H70a} uses a special class of Markov chains which obey detailed balance \eq{db} by construction. The basic idea is to break the calculation of the transition probability $x\rightarrow y$ in two steps. First, a transition from $x$ to $y\neq x$ is proposed with probability $T_{yx}$. Then, this transition is accepted with probability $A_{yx}$ and otherwise rejected, in which case, the state remains $x$. The overall transition probability is thus
\begin{align}
\cW_{yx} =  \left\{
\begin{array}{ll} 
T_{yx}A_{yx} & {\rm if}\ y \neq x \\
1-\sum_y T_{yx}A_{yx}& {\rm if}\ y=x.
\label{eq:prob}
\end{array}
\right.
\end{align}
The detailed balance condition \eq{db} becomes
\begin{equation}
R_{xy} := \frac{A_{yx}}{A_{xy}} = \frac{\pi_y}{\pi_x} \frac{T_{xy}}{T_{yx}},
\end{equation}
which in the Metropolis-Hastings algorithm is solved with the choice
\begin{equation}
A_{yx} = \min\left(1,  R_{xy}\right).
\end{equation}
We note that our quantum algorithm can also be applied to the Glauber, or heat-bath, choice~\cite{Glauber1963,Vucelja2016} 
\begin{equation}
A_{yx} = \frac 1{1+R_{yx}}.
\end{equation}

The Metropolis-Hastings algorithm is widely used to generate a Boltzmann distribution with applications in statistical physics and machine learning. Given a real energy function $E(x)$ on the configuration space $X$, the Boltzmann distribution at inverse temperature $\beta$ is defined as $\pi_x^\beta = \frac 1{Z(\beta)} e^{-\beta E(x)}$ where the partition function $Z(\beta)$ ensures normalization. In this setting, it is common practice to choose a symmetric proposed transition probability $T_{yx} = T_{xy}$, so the acceptance probability depends only on the energy difference 
\begin{equation}
A_{yx} = \min\left(1,  e^{\beta[E(x)-E(y)]} \right).
\end{equation}
Note that the Metropolis-Hastings algorithm can be applied to quantum mechanical Hamiltonians  \cite{TOVP11a}, where it can also benefit from a quadratic speed-up using Szegedy's quantization procedure  \cite{YA12a}.

%For concreteness, in the remaining of this Article, we will consider an Ising spin model where $X=\{+1,-1\}^n$, and the energy function is some simple function of $X = (x_1,x_2,\ldots x_n)$, for instance $E(x) = \sum_j h_j x_j + \sum_{jk} J_{jk} x_jx_k$. Moreover, we will assume that 
%\begin{equation}
%T_{xy} = \left\{
%\begin{array}{ll}
%\frac 1n & {\rm if} \quad |x-y| = 1\\
%0 & {\rm else}.
%\end{array}\right.
%\end{equation}
%In other words, the proposed transition consists of a random single-spin flip.

\section{Circuit for Walk operator}

Quantum algorithms built from quantization of classical walks \cite{Amb04a,S04a,SBBK08a,MNRS11a} usually assume an oracle formulation of the walk operator, where the ability to implement the transformation $W$ of \eq{W} is taken for granted. As we discuss below in Appendix \ref{sec:WalkOracle}, this transformation requires costly arithmetic operations. One of the key innovations of this article is to provide a detailed and simplified implementation of a walk operator along with a detailed cost analysis of Metropolis-Hastings walks. As it will become apparent, our implementation circumvents the use of $W$ altogether.

For concreteness, we will assume  a $(k,d)$-local Ising model, where $X=\{+1,-1\}^n$, and the energy function takes the simple form
\begin{equation}
E(x) = \sum_{\ell} J_\ell \prod_{s\in \Omega_\ell} x_s,
\label{eq:energy}
\end{equation}
where $\Omega_\ell$ are subsets of at most $k$ Ising spins, $J_\ell$ are real coupling constants where $\ell$ ranges over all the possible couplings (from 1 to $\frac{nd}{k}$), and each spin interacts with at most $d$ other spins. Note that for $k=2$ and $d\geq3$, finding the ground state is an NP-hard problem\cite{Barahona1982}.% In particular, we will assume that $E(x)$ is easy to compute, so we can efficiently implement the transformation $G:\ket x\otimes \ket 0 \rightarrow \ket x\otimes \ket{E(x)}$. 

As it is always the case for Ising models, we will assume that the proposed transitions of the Metropolis-Hastings walk are obtained by choosing a random set of spins and inverting their signs. In other words, $T_{yx} = f(x\cdot y)$ where the product is taken bit by bit and where $f(z)$ is some simple probability distribution on $X-\{1^n\}$ (it does not contain a trivial move), so $T_{yx}$  is clearly symmetric. The distribution $f(z)$ is sparse, in the sense that it has only $N\in O(n)$ non-zero entries. {}

For concreteness, we will suppose that $f$ is uniform over some set $\cM$ of moves, with $|\cM| = N$:
\begin{equation}
T_{xy} = \left\{
\begin{array}{ll}
\frac 1N & {\rm if}\ z = x\cdot y \in \cM \\
0 & {\rm otherwise}
\end{array}\right. .
\end{equation}
The most common example consists of single-spin moves, where a single spin is chosen uniformly at random to be flipped. More generally, we  will suppose that moves are sparse in the sense that each move $z_j\in \cM$ flips a constant-bounded number of spins and that each spin belongs to a constant-bounded number of different moves. For $j=1,2,\ldots N$, we use $f(j)$ as a shorthand for $f(z_j)$. With a further abuse of notation, we view $z_j\in \cM$ both as Ising spin configurations and as subsets of $[n]$, where the correspondence is given by the locations of $-1$ spins in $z_j$.

A direct implementation of the unitary $W$ generally requires costly quantum circuits involving arithmetic operations. The complexity arises from the need to uncompute a move register and a Boltzmann coin when implementing $W$. This turns out to be non-trivial and costly if a move is rejected.
Consequently, we do not implement $W$, but instead present a circuit which is isometric to the entire walk operator $U_\cW$, thus avoiding the problem. In other words, we  construct a circuit  for $\tilde U_\cW := Y^\dagger U_\cW Y$ where $Y$ maps
\begin{align}
Y: \ket x\otimes \ket y \rightarrow \left\{
\begin{array}{ll}
\ket x \otimes \ket{x\cdot y} \ & {\rm if} \ x\cdot y \in \cM\\
0 & {\rm otherwise}.
\end{array}
\right.
\end{align}
To minimize circuit depth, the second register above is encoded in a unary representation, so it contains $N$ qubits and $\ket z$ is encoded as $\ket {00\ldots 0100\ldots}$ with a $1$ at the $z$-th position. Since the state is already encoded in $N$ qubits, unary encoding adds only a small multiplicative number of qubits compared to binary encoding. In addition to these two registers, the circuit acts on an additional coin qubit. Thus, we will denote the System, Move, and Coin registers with corresponding subscripts  $\ket x_S \ket z_M \ket b_C$, and they contain $n$, $N$, and $1$ qubits respectively.  

Our implementation of the walk operator combines four components:
\begin{equation}
\tilde U_\cW =  R V^\dagger B^\dagger FBV
\end{equation} 
where 
\begin{align}
V :&\ \ket 0_M \rightarrow \sum_j \sqrt{f(j)}\ket j_M, \label{eq:move}\\
B :&\ \ket x_S \ket j_M \ket 0_C \nonumber \\
&\ \rightarrow \ket x_S \ket j_M  \left (\sqrt{1-A_{x\cdot z_j,x}}\ket 0 + \sqrt{A_{x\cdot z_j,x}}\ket 1\right)_C, \label{eq:coin}\\
F :&\ \ket x_S\ket j_M\ket b_C \rightarrow \ket{x\cdot z_j^b}_S\ket j_M \ket b_C, \ {\rm and} \label{eq:flip} \\
R :&\ \ket 0_M\ket 0_C \rightarrow -\ket 0_M \ket 0_C, \nonumber \\
&\  \ket j_M \ket b_C \rightarrow \ket j_M \ket b_C \ {\rm for}\ (j,b) \neq (0,0) \label{eq:reflection}
\end{align}
While these definitions differ slightly from the ones of Sec. \ref{sec:walk}, it can be verified straightforwardly that these realize the desired walk operator, similar to our discussion in Sec. \ref{sec:walk}. In what follows, we provide a complete description of each of these components, and their complexity is summarized in Table~\ref{cost}.

\begin{table*}
\centering
\begin{tabular}{|c||c|c|c|c|}
\hline
{\bf Gate} & {\bf 3L depth} & {\bf 3L count} & {\bf Total depth} & {\bf Qubits} \\
\hline \hline
$V$ & $\log_2 N+1$ & $2N$ & $\log_2 N+1$ & $2N$\\
$F$ & $1$ & $N$ & $\log_2 N+1$ & $2N+n$ \\
$R$ & $2\log N$ & $4N$ & $2\log N$ & $2N$ \\
$B$ & $\cO(2^d \log\frac 1\epsilon)$ & $\cO(N 2^d \log\frac 1\epsilon)$ & $\cO(\log N 2^d \log\frac 1\epsilon)$ & $2N+n+2$ \\
\hline
\end{tabular}
\caption{Upper bound on the complexity of each component of the walk operator. The cost is measured in terms of number of gates in the 3rd level of the Clifford hierarchy, which is equivalent to $T$ depth up to a small multiplicative factor. These are evaluated for a $(k,d)$-local Ising model with moves consisting of single-spin flips, in which case $N=n$. These costs could otherwise increase by a constant multiplicative amount determined by $k$, $d$ and the sparsity of the moves $z\in \cM$.} 
\label{cost}
\end{table*} 

\subsection{Move preparation $V$} 

Recall that the Move register is encoded in unary. For a general distribution $f$, the method of \cite{RG02a} can be adapted to realize the transformation \eq{move}. Here, we focus on the case of a uniform distribution. 

To begin, suppose that $N$ is a power of 2. Starting in the state $\ket{000\ldots 01}_M$, the state $\frac 1N \sum_j \ket j_M$ (in unary) is obtained by applying a sequence of $N$ gates $\sqrt{\rm SWAP}$ in a binary-tree fashion. To see this, recall that $\sqrt{\rm SWAP}\ket{10} = \frac1{\sqrt 2}(\ket{01} + \ket{\rm 10})$. The gate  $\sqrt{\rm SWAP}$ is in the third level of the Clifford hierarchy, so it can be implemented exactly using a constant number of $T$ gates.  This represents a substantial savings compared to the method of \cite{RG02a} for a general distributions which requires arbitrary rotations obtained from costly gate synthesis.

When $N$ is not a power of 2, in order to avoid costly rotations, we choose to pad the distribution with additional states and prepare a distribution $\frac 1{2^\ell} \sum_j^{2^\ell} \ket j_M$ where $\ell = \lceil \log_2 N\rceil$. The states $j=1,2,\ldots N$ encode the $N$ moves $\cM$ of the classical walk $x\rightarrow y=x\cdot z_j$, while the additional states $j>N$ correspond to trivial moves $x\rightarrow x$. This padding has the effect of slowing down the classical walk by a factor $2^\ell/N < 2$, and hence the quantum walk by a factor less than $\sqrt 2$, which is less than the additional cost of preparing a uniform distribution over a range which is not a power of $2$.

\subsection{Spin flip $F$}

The operator $F$ of \eq{flip} flips a set of system spins $z_j$ conditioned on the coin qubit and on the $j$-th qubit of the move register being in state 1. This can be implemented with at most $Nc$ Toffoli gates (controlled-controlled-NOT), where the constant $c$ upper-bounds the number of spins that are flipped by a single move of $\cM$. The coin register acts as one control for each gate, the $j$-th bit of the move register acts as the other control, and the targets are the system register qubits that are in $z_j$, for $j=1,2,\ldots N$. No gate is applied to the padding qubits $j>N$. 

This implementation has the disadvantage of  being purely sequential. An alternative implementation uses $\cO(N)$ additional scratchpad qubits but is entirely parallel. The details of the implementation depends on the sparsity of the moves $\cM$, and in general there is a tradeoff between the scratchpad size and the circuit depth. When the moves consist of single-spin flips for instance, this uses $N$ CNOTs in a binary-tree fashion (depth $\log_2 N$) to make $N$ copies of the coin qubit. The Toffoli gates can then be applied in parallel for each move, and lastly the CNOTs are undone.

\subsection{Reflection $R$}

The transformation $R$ of \eq{reflection} is a reflection about the state $\ket{00\ldots 0}_M \ket 0_C$. Using standard phase kickback methods, it can be implemented with a single additional qubit in state $\frac 1{\sqrt 2}(\ket 0 - \ket 1)$ and an open-control${}^{(N+1)}$-NOT gate. The latter can be realized from $4(N-1)$ serial Toffoli gates \cite{BBC+95a} and linear depth. 

Since our goal is to minimize circuit depth, we use a different circuit layout that uses at most $N$ ancillary qubits and $4N$ Toffoli gates to realize the $(N+1)$-fold controlled-not. The circuit once again proceeds in a binary tree fashion, dividing the set of $N+1$ qubits into $(N+1)/2$ pairs and applying a Toffoli gate between every pair with a fresh ancilla in state 0 as the target. The ancillary qubit associated to a given pair is in state $0$ if and only if both qubits of the pair are in state 0. The procedure is repeated for the $(N+1)/2$ ancillary qubits, until a single bit indicates if all qubits are in state 0. The ancillary bits are then uncomputed. Thus, the total depth in terms of gates in 3rd level of the Clifford hierarchy is $2\log_2N$.

\subsection{Boltzmann coin $B$} 

The Boltzmann coin given in \eq{coin} is the most expensive component of the algorithm, simply because it is the only component which requires rotations by arbitrary angles. Specifically, conditioned on move qubit $j$ being 1 and the system register being in state $x$, the coin register undergoes a rotation by an angle 
\begin{equation}
\theta_{x,j} =  \arcsin \left(\sqrt{\min\{e^{-\beta\Delta_j(x)},1\}}\right)
\label{eq:arcsin}
\end{equation}
for Metropolis-Hastings or 
\begin{equation}
\theta_{x,j} =  \arcsin \left(\frac 1{1+e^{\beta\Delta_j(x)}}\right)
\label{eq:arcsinG}
\end{equation}
for Glauber dynamics, 
where $\Delta_j = E(x\cdot z_j)-E(x)$. Given the sparsity constraints of the function $E$ and of the moves $z_j\in \cM$, the quantity $\Delta_j$ can actually be evaluated from a subset of qubits of the system register, namely $\cN_j = \{k | k \in \Omega_\ell , \ z_j\cap \Omega_\ell \neq \emptyset,\ \forall \ell\}$.  For single-spin flips on a $(k,d)$-local Hamiltonian, $|\cN_j| \leq kd$ by definition. For  multi-spin flips $z_j$, we get $|\cN_j| \leq |z_j| kd$. 

Thus, the Boltzmann coin consists of a sequence of $N$ conditional gates $R_j$, where $R_j$ itself is a single-qubit rotation by an angle determined by the qubits in the set $\cN_j$. Since each $\cN_j$ is of constant-bounded size, each $R_j$ can be realized from a constant number of $T$ gates, so the entire Boltzmann coin requires $\cO(N \log \frac 1\epsilon)$ $T$ gates, where $\epsilon$ is the desired accuracy for the synthesis of single-qubits rotations. It is likely that a high precision is needed to ensure the detailed balance condition. We leave for future research the numerical investigation of how low the precision can be without causing significant errors.  

Because all gates $R_j$ act on the Coin register, they must be applied sequentially. An alternative consists in copying the Coin register in the conjugate basis of $\sigma_y$, i.e. $\ket{\pm i} \rightarrow \ket{\pm i}^{\otimes N}$ since a sequence of rotations $e^{i\theta_j\sigma_x}$ is equivalent to a tensor product of these rotations under this mapping. Moreover, any set of gates $R_j$ with non-overlapping $\cN_j$ can be executed in parallel. Consequently, the total depth can be bounded by a constant at the expense of $N$ additional qubits.

The complexity of the Boltzmann coin does scale exponentially with the sparsity parameters of the model however, namely as $\cO(\max_j 2^{|\cN_j|})$. A circuit that achieves $R_j$ consists of a sequence of $2^{|\cN_j|}$ single-qubit rotations by an angle given by \eq{arcsin} or \eq{arcsinG}, conditioned on the bits in $\cN_j$ taking some fixed value. Each of these $2^{|\cN_j|}$ multi-controlled rotations require $\cO(|\cN_j|)$ Toffoli gates along with $\cO(\log\frac 1\epsilon)$ $T$ gates, for an overall circuit depth of $\cO(2^{|\cN_j|}|\cN_j|\log\frac 1\epsilon)$ to realize $R_j$.  

Perhaps a more efficient way to realize the Boltzmann coin uses  quantum signal processing methods \cite{LYC16a,LC16a,LC17a,H18a}. This is a method to construct a unitary transformation $S_2 = \sum_x f(e^{i\phi_x})\kb{x}{x}$ from a controlled version of $S_1 = \sum_x e^{i\phi_x}\kb{x}{x}$. In the current setting, $S_1 = \sum_x e^{\lambda i\Delta_j(x)}\kb xx$ and we choose $f(e^{i\lambda\Delta_j(x)}) = e^{i2\theta_{x,j}}$ where $\theta_{x,j}$ is given at \eq{arcsin}. Applying a Hadamard to the Coin qubit, followed by a controlled $S_2$ with the Coin acting as control, and followed by a Hadamard on the Coin qubit again results in the transformation that we called $R_j$ above and that builds up the Boltzmann coin transformation $B$.

Above, the constant $\lambda$ is chosen in such a way that the argument of the exponential $e^{i\lambda\Delta_j(x)}$ is restricted to some finite interval which does not span the entire unit circle, say in the range $[-\pi/2,\pi/2]$. The exponential can be further decomposed as a product 
\begin{equation}
e^{i\lambda \Delta_j(x)} = \prod_{\ell: \Omega_\ell \cap  z_j \neq \emptyset} \exp\Big\{i\lambda 2J_\ell \prod_{s\in \Omega_\ell} x_s\Big\}.
\label{eq:random_ising}
\end{equation}
Each of these factors is a rotation by an angle $2J_\ell$, whose sign is conditioned on the parity of the bits in $\lambda\Omega_\ell$. The parity bit can be computed using $|\Omega_\ell|$ CNOTs, and the rotation is implemented using gate synthesis, with a $T$-gate count per transformation of $\cO(\log \frac 1\epsilon)$, which is dictated by the accuracy $\epsilon$. 
The complexity of quantum signal processing depends on the targeted accuracy. More precisely, it scales with the number of Fourier coefficients required to approximate the function $f(e^{i\theta}) = \min(1,e^{-\theta\beta/\lambda})$ or $g(e^{i\theta}) = \frac 1{1+e^{\theta\beta/\lambda}}$  to some constant accuracy $\epsilon$ on the domain $\theta \in [-\pi/2,\pi/2]$. 

Quantum signal processing, or alternative methods, will offer an advantage on some models, when there are different couplings and a high number of body interactions for example. The scaling of these methods is case dependent. Indeed, it will highly depend on these couplings and the number of spin flips $z_j$.  

\section{Heuristic use}

The Metropolis-Hastings algorithm is widely used heuristically to solve minimization problems using simulated annealing or related algorithms \cite{KGV83a}. The objective function is the energy $E(x)$. Starting from a random configuration or an informed guess, the random walk is applied until some low-energy configuration $x$ is reached. The parameter $\beta$ can be varied in time, with an initial low value enabling large energy fluctuations to prevent the algorithm from getting trapped in local minimums, and large final value to reach a good (perhaps local) minimum. 

In this section, we propose heuristic ways to use the quantum walk in the context of a minimization problem. We first recall the concept of total time to solution \cite{RWJB14a} which we use to benchmark and compare different heuristics. We then present two quantum heuristics which we compare  using numerical simulations on small instances.

Since the purpose of our study is to compare a classical walk to its different quantum incarnations -- as opposed to optimizing a classical walk -- we will use a schedule with a linearly increasing value of $\beta$ in time up to a fixed final value of $\beta$  in our comparison and expect our conclusions to hold if an optimized $\beta$ schedule was used instead in both the classical and the quantum walks.

\subsection{Total time to solution}

When a random walk is used to minimize some function $E(x)$, the minimum $x^*$ is only reached with some finite probability $p$. Starting from some distribution $q(x)$ and applying the walk $\cW$ sequentially $t$ times, the success probability is $p(t) = (\cW^t q)(x^*)$.  To boost this probability to some constant value $1-\delta$, it is sufficient to repeat the procedure $L = \frac{\log(1-\delta)}{\log(1-p(t))}$ times. The total time to solution is then defined as the duration of the walk $t$ times the number of repetitions $L$,
\begin{equation}
{\rm TTS}(t) := t \frac{\log(1-\delta)}{\log(1-p(t))}.
\end{equation}

 There is  a compromise to be reached between the duration of the walk $t$ and the success probability $p(t)$ -- longer walks can reach a higher success probability and therefore be repeated fewer times, but increasing the duration $t$ of the walk beyond a certain point has a negligible impact on its success probability $p(t)$. We thus define the minimum total time to solution as min(TTS) $= \min_t{\rm TTS}(t)$.

\subsection{Zeno with rewind}

In \Sec{zeno}, we explained how to prepare the eigenstate of $U_\cW$ with eigenvalue 1 using a sequence of walks $\cW^0, \cW^1,\ldots, \cW^L=\cW$. In the setting of Metropolis-Hastings where $\cW$ is the walk with parameter $\beta$, a natural choice of $\cW^j$ is given by $\beta^j = \frac jL\beta$. An optimized $\beta$ schedule is also possible, but for a systematic comparison with the classical walk, we choose this fixed schedule, whose only parameter is the number of steps $L$.

Let us revisit the argument of \Sec{zeno} to establish some notation. Define the binary projective measurement $\{Q_j,Q_j^\perp\} := \{\kb{\pi^j}{\pi^j}, I- \kb{\pi^j}{\pi^j}\}$. This binary measurement can be realized from $\frac 1{\delta_j}$ uses of $U_{\cW^j}$, where $\delta_j$ denotes the spectral gap of $U_{\cW^j}$. Starting from the state $\ket{\pi^0}$, the Zeno algorithm consists in performing the sequence  of binary measurements $\{Q_j,Q_j^\perp\}$ in increasing value of $j$. The outcome $Q_j$ on state $\ket{\pi^{j-1}}$ yields state $\ket{\pi^j}$ and occurs with probability $F_j^2 := |\bracket{\pi^{j-1}}{\pi^j}|^2$. The sequence of measurements  succeeds if they all  yield this outcome, which occurs with probability $\prod_{j=1}^L F_j^2$ and requires $\sum_{j=1}^L \frac 1{\delta_j}$ applications of quantum walk operator. For the algorithm to be successful, the final measurement in the computational basis must also yield the optimal outcome $x^*$, which occurs with probability $\pi^L(x^*)$. Thus, the total time to solution for an $L$-step algorithm is  
\begin{equation}
{\rm TTS}(L) = \frac{\log(1-\delta)}{\log(1-\pi^L(x^*)\prod_{j=1}^L F_j^2 )} \sum_{j=1}^L \frac 1{\delta_j}.
\end{equation}

In the method outlined above, a measurement outcome $Q_j^\perp$ requires a complete restart of the algorithm to $\beta = 0$. There  exists an alternative to a complete restart  which we call rewind. It was first described in the context of Zeno state preparation in Ref. \cite{LP19a}, but originates from Refs. \cite{MW05a,TOVP11a}. It consists of iterating between the measurements $\{Q_{j-1},Q_{j-1}^\perp\}$ and $\{Q_j,Q_j^\perp\}$ until the measurement $Q_j$ is obtained. It can easily be shown that a transition between outcomes $Q_{j-1}\leftrightarrow Q_j$ or $Q_{j-1}^\perp\leftrightarrow Q_j^\perp$ is $F_j^2$ while the probability of a transition between outcomes $Q_{j-1}\perp \leftrightarrow Q_j$ or $Q_{j-1}\leftrightarrow Q_j^\perp$ is $1-F_j^2$. Given the cost $\frac 1{\delta_j}$ of each of these measurements, we obtain a simple recursion relation for the expected cost of a successful $\ket{\pi^{j-1}} \rightarrow \ket{\pi^j}$ transition with rewind, and thus for the total time to solution for a $L$-step Zeno protocol with rewind. The minimal total time to solution is obtained by minimizing over $L$.  In Ref. \cite{LP19a}, it was found that rewinding yields substantial savings compared to the regular Zeno strategy for the preparation of quantum many-body ground states.

\subsection{Unitary implementation}

We propose another heuristic use of the quantum walk which does not use measurement. Starting from state $\ket{\pi^0}$, it consists in  applying the quantum walk operators $U_{\cW^j}$ sequentially, resulting in the state
\begin{equation}
\ket{\psi(L)} = U_{\cW^L}\ldots U_{\cW^2}U_{\cW^1}\ket{\pi^0},
\end{equation} 
and ending with a computational basis measurement.
The algorithm is successful if a computational basis measurement yields the outcome $x^*$ on state $\ket{\psi(L)}$ (rewind could be used otherwise), so the total time to solution for the $L$-step algorithm is 
\begin{equation}
{\rm TTS}(L) = \frac{\log(1-\delta)}{\log(1-|\bra{x^*}U_{\cW^L}\ldots U_{\cW^2}U_{\cW^1}\ket{\pi^0}|^2)}.
\end{equation}

While we do not have a solid justification for this heuristic use, in Ref. \cite{BKS09a}, a protocol was proposed which used a similar sequence of unitaries, but where each unitary was applied a random number of times. The motivation for these randomized transformations was to phase randomize in the eigenbasis of the instantaneous unitary operator. When the spectral gap of a unitary operator is $\delta$ and that unitary is applied a random number of times  in the interval $[0,\frac 1{\delta_j}]$, then the relative phase between the eigenstate with eigenvalue 1 and the other eigenstates is randomized over the unit circle, thus mimicking the effect of a measurement (but with an unknown outcome). From this analogy, we could expect that the unitary implementation yields a minimal total time to solution roughly equal to the Zeno-based algorithm with no rewind. But as we will see in the next section, its behavior is much better than anticipated -- this method is more efficient than the Zeno algorithm with rewind, which itself is more efficient than Zeno without rewind. 

\subsection{Numerical results}

\begin{figure}[t!]
\includegraphics[width=\columnwidth]{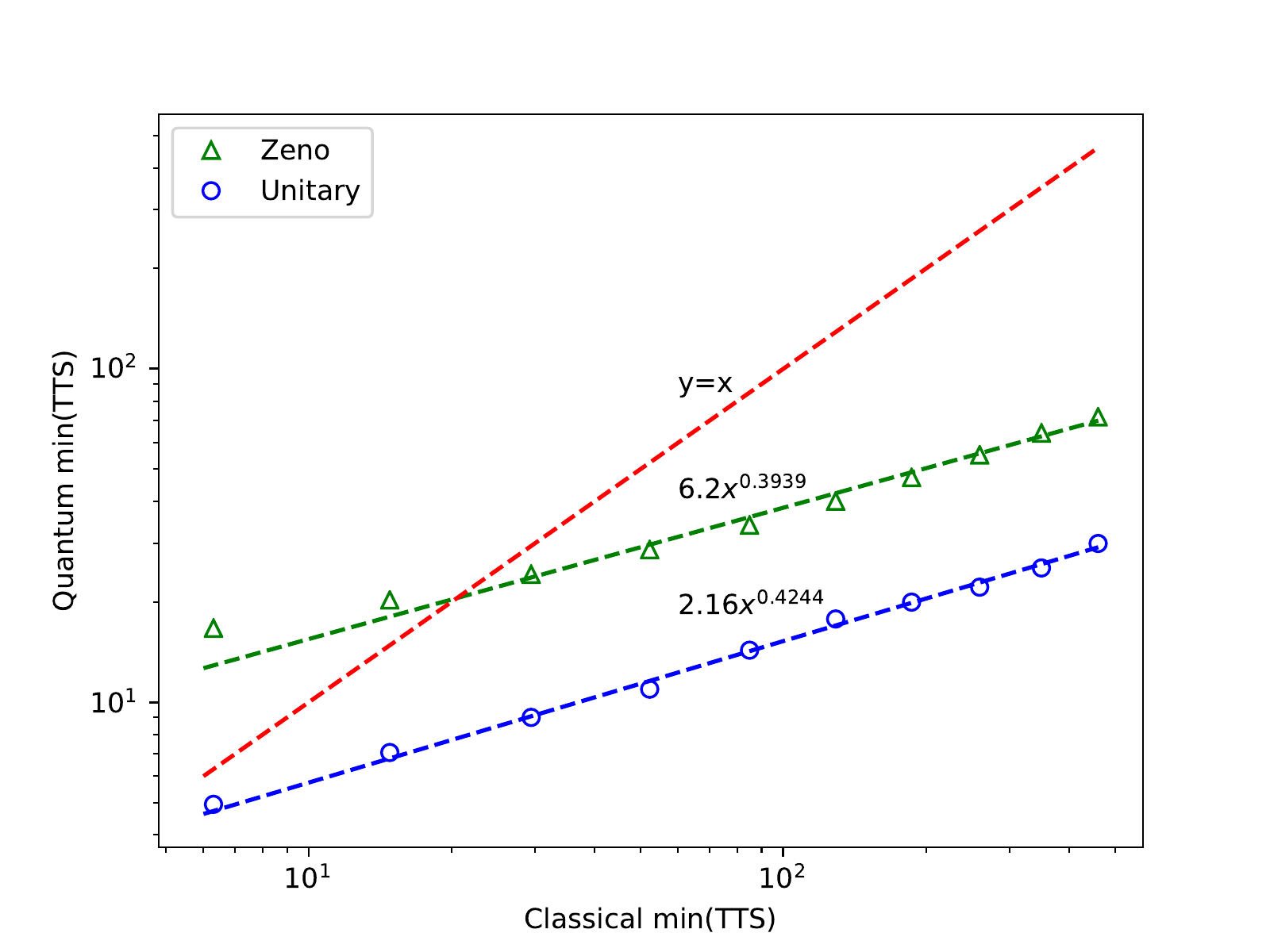}
\caption{Quantum versus classical minimum total time to solution (min(TTS))  for a one dimensional Ising model of length ranging from $n=3$ to 12 at $\beta = 2$.  The line $x=y$ is shown for reference of a quantum speedup.}
\label{fig:TTS_1D}
\end{figure}

We have numerically benchmarked three heuristic algorithms: the classical walk with a variable-length linear interpolation between $\beta=0$ and $\beta=2$ and starting from a uniform distribution; the discrete, or Zeno-based adiabatic algorithm with rewind; and the unitary algorithm of the last subsection. The first system considered is a one dimensional Ising model.  Figure \ref{fig:TTS_1D} shows the quantum versus classical minimal total time to solution. The results clearly indicate a polynomial advantage of the quantum algorithms over the classical algorithm. Surprisingly, both quantum approaches show a similar improvement over the classical approach that exceed the expected quadratic speedup, with a power law fit of 0.42 using the unitary algorithm and 0.39 using the Zeno algorithm.

\begin{figure}[t!]
\includegraphics[width=\columnwidth]{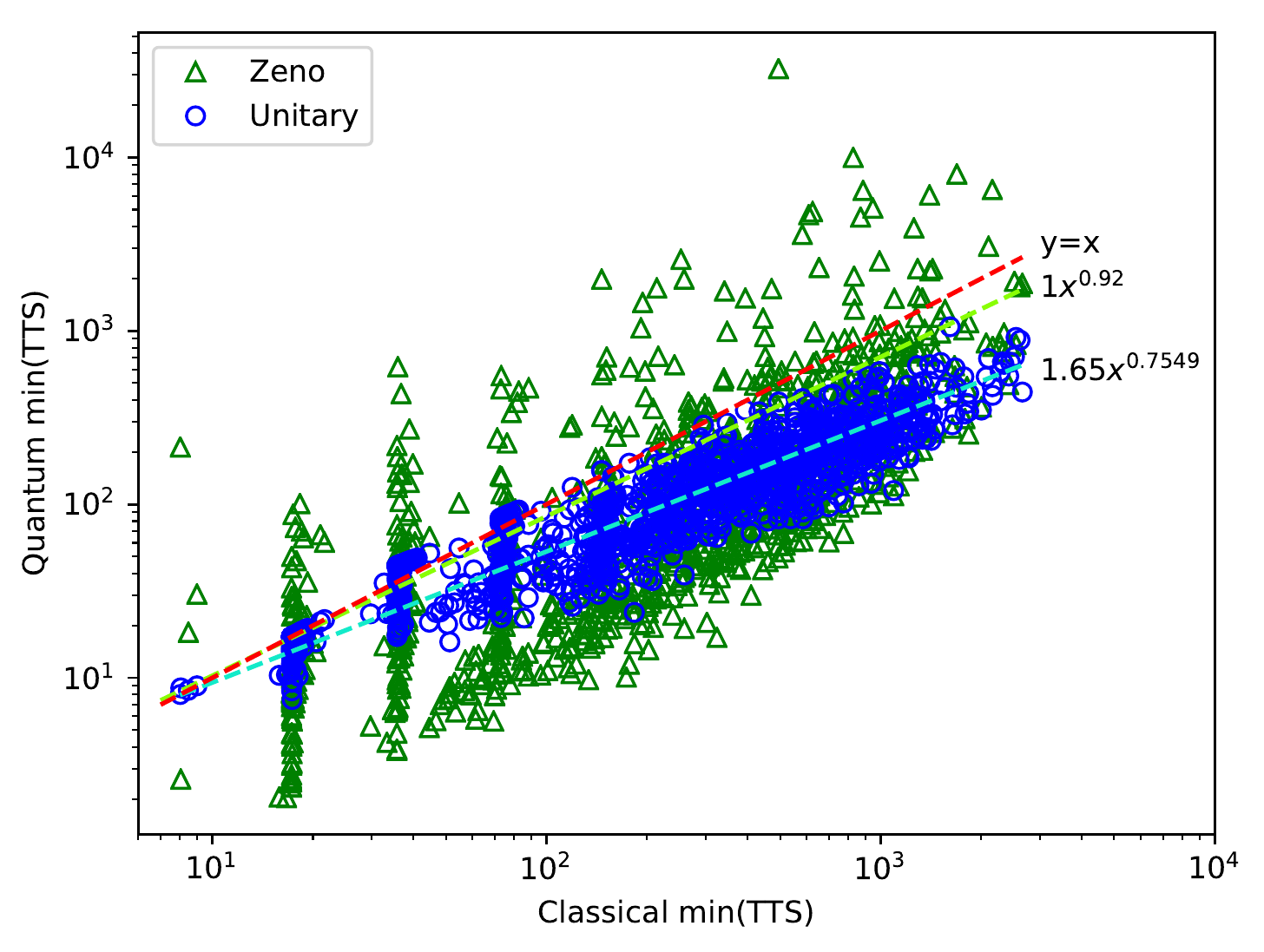}
\caption{Quantum versus classical minimum total time to solution (min(TTS))  for a random sparse Ising model at $\beta = 2$, $k=2$ and $d=3.5n$. 100 random problem instances are chosen for each size, ranging from $n=4$ to 14 .}
\label{fig:TTS}
\end{figure}

The second system considered is a sparse random Ising model: it has  gaussian random couplings $J_\ell $ of variance 1, and the interactions sets $\Omega_\ell$ (c.f. \eq{random_ising}) consist of a random subset of $3.5n$ of all the $n(n-1)/2$ pairs of sites.
Figure \ref{fig:TTS} shows quantum versus classical minimum total time to solution  for a random ensemble of 100 systems of each sizes $n=4$ to 14. We observe that the unitary algorithm is consistently faster than the classical algorithm, with an average polynomial speedup of degree 0.75, less than the expected quadratic gain. Moreover, the different problem instances are all quite clustered around this average behavior, suggesting that the quantum speedup is fairly general and consistent.  In contrast the Zeno algorithm shows large fluctuations about its average, particularly on very small problem instances. The average polynomial speedup is of degree 0.92, far worse than the unitary algorithm. Overall, the results indicate a polynomial advantage of the quantum methods over the classical method, but these advantages are much less pronounced than for the 1D Ising model. 

In both the one-dimensional and the random graph Ising model, the unitary quantum algorithm achieves very similar and sometimes  superior scaling to the Zeno with rewind algorithm. This is surprising given the observed improvement obtained from rewind in Ref. \cite{LP19a} and our expectation that the unitary algorithm behaves essentially like Zeno without rewind.

%\begin{table}
%\begin{tabular}{|l|ll|}
%\hline
%Method & $a$ & $b$ \\
%\hline
%Classical median & $0.17\pm 0.01 $ & $3.36 \pm 0.04$ \\
%Unitary median & $0.9\pm 0.1 $ & $2.18 \pm 0.07$ \\
%Zeno median & $0.3\pm 0.1 $ & $2.62 \pm 0.2$ \\
%Classical 10 centile & $0.053\pm 0.005 $ & $4.11 \pm 0.04$ \\
%Unitary 10 centile & $0.57\pm 0.07 $ & $2.72 \pm 0.07$ \\
%Zeno 10 centile & $1\pm 2 $ & $2.84 \pm 0.7$ \\
%\hline
%\end{tabular}
%\caption{Fit parameters $a$ and $b$   used in \fig{TTS} of the form min(TTS) = $an^b$. Error bars are obtained from the bootstrap method.}
%\label{fit}
%\end{table}

\section{Discussion}

Our conclusion, and perhaps one of the key messages of this Article, is that even though the quantum walk is traditionally defined with the help of a walk oracle, its circuit implementation does not necessarily require it, and this can lead to substantial savings. In Appendix \ref{sec:WalkOracle}, we discuss the difficulty of implementing the quantum walk unitary $W$.  Appendix \ref{irreversible} presents an improved parallelized heuristic classical walk for discrete sparse optimization problems which could potentially lead to significant improvements on a quantum computer. Unfortunately this walk is not reversible, which motivates further generalization of Szegedy's quantization to include  irreversible classical walks. In the rest of this section, we discuss the prospect of using the quantum walk to outperform a classical supercomputer.

We have  proposed heuristic quantum algorithms based on the Szegedy walk for solving discrete optimization problems. Theoretical bounds show that the quantum algorithm can benefit from a quadratic speed-up ($x^{0.5}$) over its classical counterpart. Our numerical simulations on small problem instances indicate a super-quadratic speed-up ($\approx x^{0.42}$) for the Ising chain, see figure~\ref{fig:TTS_1D}, and sub-quadratic speed-up ($\approx  x^{0.75}$) for random sparse Ising graphs, see figure~\ref{fig:TTS}. It remains an interesting question to understand more broadly what type of problems can benefit from what range of speed-up and why. With these crude estimates in hand we can already look into the achievability of a quantum  speed-up on realistic devices.  

We will compare performances to the special-purpose supercomputer ``Janus'' \cite{C09b,C12a} which consists of a massive parallel field-programmable gate array (FPGA). This system is capable of performing $10^{12}$ Markov chain spin updates per second on a three-dimensional Ising spin glass of size $n=80^3$.  A calculation that lasts a bit less than a month will thus realize $10^{18}$ Monte Carlo steps. On the one hand, assuming that the theoretically predicted quadratic speed-up holds and since the numerics show a constant factor around $1$, the quantum computer must realize at least $10^9$ steps per month in order to keep up with the classical computer.  This requires that a single step of the quantum walk be realized in a few milliseconds. On the other hand, the super-quadratic speed-up we have observed would allow almost a tenth of a second to realize a single quantum step, while the sub-quadratic speed-up would require that a single step be realized within 0.1 microseconds. 

Taking the circuit depth reported in Table \ref{cost} as reference with $d=6$ for a three-dimensional lattice leads to a circuit depth of $\log(80^3)\times 2^6 \approx 1000$.  To avoid harmful error accumulation, the gate synthesis accuracy $\epsilon$ should be chosen as the inverse volume (circuit depth times the number of qubits) of the quantum circuit, roughly $\epsilon^{-1} \approx 80^3 \times \log(80^3)\times 10^9 \approx 10^{16}$,  so on the order of $4 \log\frac 1\epsilon \approx 200$ logical $T$ gates are required per fine-tuned rotation \cite{RS16a,BRS15a}, for a total logical circuit depth of 200,000. With these estimates, the three scenarios described above require logical gate speeds ranging from an unrealistically short 0.5 picoseconds (sub-quadratic speed-up), to an extremely challenging 1 nanosecond (quadratic speed-up), and allow 0.5 microseconds (super-quadratic speed-up). 

We could instead compile the rotations offline and teleport them in the computation \cite{JWM+12}, which requires at least $4 \log\frac 1\epsilon \approx 200$  more qubits, but increases the time available for a logical gate by the same factor. Under this scenario, the time required for each logical gate would range from 0.1 nanoseconds (sub-quadratic speed-up), to 20 microseconds (quadratic speed-up), and to 1 miliseconds (super-quadratic speed-up). These estimates are summarized in Table \ref{table:speed}. The latter is a realistic logical gate time for many qubit architectures, while there is no current path to achieve nanosecond logical gate times.

Given the above analysis, if a quantum computer is to offer a practical speed-up, we conclude that a better understanding of the class of problems for which heuristic super-quadratic speed-ups can be achieved is required,  and that we need to optimize circuit implementations even further. 

\begin{table*}
\centering
\begin{tabular}{|l||l|l|}
\hline
{\bf Quantum speedup} & {\bf Synthesis online}  & {\bf Synthesis offline} \\
\hline \hline
Sub-quadratic $x^{0.75}$ & 0.5ps & 0.1ns\\
Quadratic $x^{0.5}$ &  1ns & 20$\mu$s \\
Super-quadratic $x^{0.42}$ & 0.5$\mu$s & 1ms \\
\hline
\end{tabular}
\caption{Logical gate time required to outperform a supercomputer capable of realizing $10^{12}$ Monte Carlo updates per nanosecond in a computation that lasts one month. Arbitrary single-qubit rotations can be synthesized online or offline at an additional qubit cost.}
\label{table:speed}
\end{table*}

\section{Acknowledgements}

We thank Jeongwan Haah, Thomas H\"aner, Matt Hastings, Guang Hao Low and Guillaume Duclos-Cianci for stimulating discussions. JL acknowledges support from the FRQNT programs of scholarships.

\bibliographystyle{plainnat}
\bibliography{JPHST20.bib}
%*======================================================*

\onecolumn\newpage
\appendix

\section{Walk oracle}
\label{sec:WalkOracle}

Our implementation of the walk operator does not make use of the walk unitary $W$ of \eq{W}. Since the transition matrix elements $\cW_{xy}$ can be computed efficiently, we know that $W$ can be implemented in polynomial time. But this requires costly arithmetics which would yield a substantially larger complexity than the approach presented above. 

To see how this complexity arises, consider the following implementation of $W$, which uses much of the same elements as introduced above. The computer comprises two copies of the system register, which we now label Left and Right. As before, it also comprises a Move register and a Coin register. Begin with the Left register in state $x$ and all other registers in state $0$. Use the transformation $V$ to prepare the state  $\ket x_L\otimes \ket 0_R \otimes \ket f_M \otimes \ket 0_C$. Using $n$ CNOTs, copy the state of the Left register onto the Right register, resulting in $\ket x_L\otimes \ket x_R \otimes \ket f_M \otimes \ket 0_C$. Apply the move $z_j$ proposed by the Move register to the Right register. If the Move register is encoded in unary representation as above, this requires $\cO(N)$ CNOTs, and results in the state
\begin{align}
&\ket x_L\otimes  \sum_{j\in \cM} \sqrt{f(z_j)} \ket{x\cdot z_j}_R\otimes \ket{z_j}_M\otimes\ket 0_C .
\label{eq:MR}
\end{align}
Using a version of the Boltzmann coin transformation on the Left, Right and Coin register yields
\begin{align}
&\ket x_L  \sum_j \sqrt{f(z_j)} \ket{x\cdot z_j}_R \ket{z_j}_M \nonumber \\
& \otimes \left(\sqrt{1-A_{(x\cdot z_j)x}}\ket 0 + A_{(x\cdot z_j)x}\ket 1\right)_C \\
&=\ket x_L  \sum_{y\neq x} \sqrt{\cW_{yz}} \ket{y}_R \ket{x\cdot y}_M \nonumber \\
&\otimes  \left(\sqrt{A_{yx}^{-1}-1}\ket 0 + \ket 1\right)_C.
\end{align}
At this point, we swap the Left and Right registers conditioned on the Coin qubit being in state 1, resulting in the state  
\begin{align}
&\sum_{y\neq x}  \sqrt{\cW_{yx}} \ket y_L  \ket{x}_R \ket{x\cdot y}_M \ket 1_C \nonumber\\
+ &\sum_{y\neq x}\sqrt{f(x\cdot y)(1-A_{yx})} \ket x_L   \ket{y}_R \ket{x\cdot y}_M \ket 0_C. \label{eq:final-state}
\end{align}
Finally, reset the move register to $0$ using $2N$ CNOTS with controls from the Left and Right registers. At this point, the move register is disentangled and discarded, resulting in the state 
\begin{align}
&\sum_{y\neq x}  \sqrt{\cW_{yx}} \ket y_L  \ket{x}_R  \ket 1_C \nonumber\\
+ &\sum_{y\neq x}\sqrt{f(x\cdot y)(1-A_{yx})} \ket x_L   \ket{y}_R \ket 0_C.
\end{align}
The relative weights of the two branches are the same as the classical MCMC methods, which corresponds to an acceptance rate of approximately $1/2$.

This is quite similar to the state that would result from the quantum walk operator $W$ of \eq{W}, save for one detail.  When the acceptance register is in state 0, the state $\sum_{y\neq x}\sqrt{f(x\cdot y)(1-A_{yx})}  \ket{y}_R$ of the right register needs to be mapped to the state $\sqrt{\cW_{xx}}\ket x_R$. Such a rotation clearly depends on all the coefficients $A_{yx}$, and all implementations we could envision used arithmetic operations that compute $A_{xy}$. 

\section{Irreversible parallel walk}
\label{irreversible}

Note that the Boltzmann operator $B$ has a total number of gates that scales with the system size $n$, even though it is used to implement a single step of the quantum walk and that on average, a single spin is modified per step of the walk.  This contrasts with the classical walk where in a single step of $\cW$, a spin transition $x \rightarrow x\cdot z$ is chosen with probability $f(z)$, the acceptance probability is computed, and the move is either accepted or rejected. Each transition $x \rightarrow x\cdot z$ typically involves only a few spins (one in the setting we are currently considering), so implementing such a transition in the classical walk does not require an extensive number of gates. The complexity in that case is actually dominated by the generation of a pseudo-random number selecting the location of the spin to be flipped.  As a consequence, the quantum algorithm suffers an $n$-fold complexity increase  compared to the quantum algorithm.

This motivates the construction of a modified classical walk for the lattice spin model which also affects every spin of the lattice, putting the classical and quantum walks on equal footing in terms of gate count. For simplicity, suppose that the set of moves  $z_i \in \cM$ consist in single-spin flips. We define a {\em parallel classical walk} with transition matrix 
\begin{equation}
\cW_{yx} = \prod_{j=1}^N \left[q B_i(x)\right]^{(1-\frac{x_i\cdot y_i}2)}  \left[1-q B_i(x)\right]^{(1+\frac{x_i\cdot y_i}2)},
\end{equation}
where $B_i(x) = \min\{1,e^{\beta[E(x)-E(x\cdot z_i)]}\}$ and $z_i$ is the transition which consists of flipping the $i$th spin only, so only spin $i$ differ in $x$ and $x\cdot z_i$. The variable $0\leq q \leq 1$ is a tunable parameter of the walk. In other words, a single step of this walk can be decomposed into a sequence over spins $i$, and consists of flipping $i$ with probability $q$ and accepting the flip with probability $B_i(x) = \min\{1,e^{\beta[E(x)-E(x\cdot z_i)]}\}$. Importantly, even if the moves are applied sequentially, the acceptance probability $B_i(x)$ is always evaluated relative to the state at the beginning of the step, even though other spins could have become flipped during the sequence.

If instead the acceptance probability was evaluated conditioned on the previously accepted moves -- i.e. $B_i(x) = \min\{1,e^{\beta[E(x\cdot z_i^{\rm tot})-E(x\cdot z_i)]}\}$ where $z_i^{\rm tot}$ is the total  transition accumulated up to step $i$ --  then this acceptance probability would be the same as used in the Metropolis-Hastings algorithm. Note that for a {\em local} spin model with, e.g., nearest-neighbor interactions, the two acceptance probabilities only differ if a neighbor of site $i$ has been flipped prior to attempting to flip spin $i$. Because a transition on each spin is proposed with probability $q$, the probability of having two neighboring spins flipped is $\cO(q^2)$. Thus, we essentially expect a single step of this modified walk to behave like $qn$ steps of the original Metropolis-Hastings walk, with a systematic error that scales like $nq^2$. Moreover, this systematic error is expected to decrease over time since once the walk settles in a low-energy configuration, very few spin transitions will turn out to be accepted, thus further decreasing the probability of a neighboring pair of spin flips. 

\begin{figure}
\centering
\includegraphics[width=0.5\columnwidth]{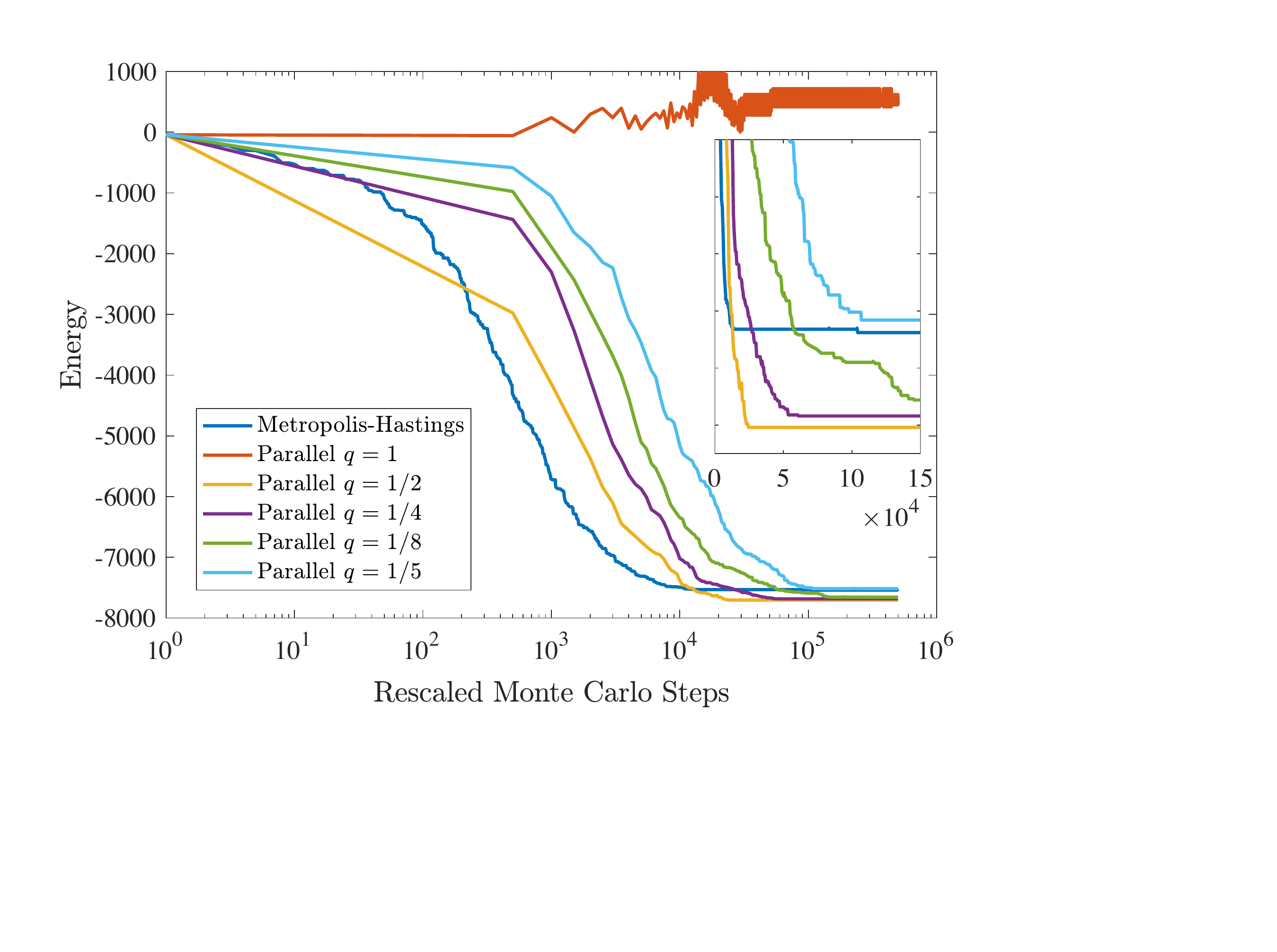}
\caption{Energy above ground state of an Ising model on a complete graph with $n=500$ vertices with random binary couplings as a function of the number of Monte Carlo steps. Results are shown for regular Metropolis-Hastings walk and the parallelized walk with different values of $q = 1,\ \frac 12,\ \frac 14,\ \frac 18$ and $\frac 1{16}$. The temperature was set to $\beta = 3$, so the fixed point should be a low energy state. Since each step of the parallelized classical walk requires $n=500$ times as many gates as the original walk, the time label of the parallel walk has been multiplied by $n$ so it adequately represent the number of computational resources. The parallel walk with $q<1$ outperforms the original walk at long times (see inset with first 150,000 steps) and achieves similar performances at short times as $q$ approaches 1. }
\label{fig:Ising}
\end{figure}

To verify the above expectation, we have performed numerical simulations on an Ising model
$$
H = \sum_{i,j } J_{i,j} x_i x_j
$$
 where $J_{i,j}$ were randomly chosen from $\{+1,-1\}$. Results are shown on \fig{Ising}. What we observe is that, for an equal amount of computational resources, the parallelized walk outperforms the original walk. This is true both in terms in reaching a quick pseudo minimum configuration at short times and in terms of reaching the true minimum at longer times. Thus, while this parallelization was introduced to ease the quantization procedure, it appears to be of interest on its own. 

In this case, the quantum walk unitary $W$ can easily be applied. We first proceed as in the previous subsection and use CNOTs to copy the Right register onto the Left register, yielding state $\ket x_L\otimes \ket x_R$. Then, sequentially over all spins $i$, apply a rotation to spin $i$ of the Left register conditioned on the state of the spin $i$ and its neighbors on the Right register. This rotation transforms $\ket{x_i} \rightarrow \sqrt{1-qB_i(x)} \ket {x_i} + \sqrt{qB_i(x)} \ket {\overline x_i}$. Note that the function $B_i(x)$ only depends on the bits of $x$ that are adjacent to site $i$, so this rotation acts on a constant number of spins so requires a constant number of gates. Thus, the cost of the classical and the quantum parallel walks have the same scaling in $n$. Combined to its observed advantages over the original classical walk, the parallel walk thus appears as the ideal version for a quantum implementation.

Unfortunately, the parallel walk is not reversible -- it does not obey the detailed-balance condition \eq{db}. Thus, it is not directly suitable to quantization \`a la Szegedy. While quantization of non-reversible walks were considered in \cite{MNRS11a}, they require an implementation of {\em time-reversed} Markov chain $\cW^*$ defined from  $\cW$ and its fixed point $\pi$ as 
\begin{equation}
\cW_{xy}^* \pi_x = \cW_{yx}\pi_x.
\end{equation}
Unfortunately, we do not know how to efficiently implement a quantum circuit for the time-reversed walk $W^*$, so at present we are unable to quantize this parallel walk.

\end{document}